# Maximum-entropy from the probability calculus: exchangeability, sufficiency

P.G.L. Porta Mana   <pgl@portamana.org>

7 June 2017

*Dedicato alla mia fantastica sorellina Marianna per il suo compleanno*

The classical maximum-entropy ~~principle~~ method (Jaynes 1963) appears in the probability calculus as an approximation of a particular *model by exchangeability* or a particular *model by sufficiency*.

The approximation from the exchangeability model can be inferred from an analysis by Jaynes (1996) and to some extent from works on entropic priors (Rodríguez 1989; 2002; Skilling 1989a; 1990). I tried to show it explicitly in a simple context (Porta Mana 2009). The approximation from the sufficiency model can be inferred from Bernardo & Smith (2000 § 4.5) and Diaconis & Freedman (1981) in combination with the Koopman-Pitman-Darmois theorem (see references in § 3).

In this note I illustrate how either approximations arises, in turn, and then give a heuristic synopsis of both. At the end I discuss some questions: Prediction or retrodiction? Which of the two models is preferable? (the exchangeable one.) How good is the maximum-entropy approximation? Is this a "derivation" of maximum-entropy?

I assume that you are familiar with: the maximum-(relative-)entropy method (Jaynes 1957a; much clearer in Jaynes 1963; Sivia 2006; Hobson et al. 1973), especially the mathematical form of its distributions and its prescription "expectations = empirical averages"; the probability calculus (Jaynes 2003; Hailperin 1996; Jeffreys 2003; Lindley 2014); the basics of models by exchangeability and sufficiency (Bernardo et al. 2000 ch. 4), although I'll try to explain the basic ideas behind them – likely you've often worked with them even if you've never heard of them under these names.

## 1   Context and notation

We have a potentially infinite set of measurements, each having $K$ possible outcomes. Dice rolls and their six outcomes are a typical example. I use the terms "measurement" and "outcome" to lend concreteness to the discussion, but the formulae below apply to much more general contexts.

The proposition that the $n$th measurement has outcome $k$ is denoted $E_k^{(n)}$. The relative frequencies of the $K$ possible outcomes in a set of measurements are denoted $\boldsymbol{f} \coloneqq (f_k)$. It may happen that in a measurement we observe not directly an outcome but an "observable" having values $(O_k) \eqqcolon \boldsymbol{O}$ for the $K$ outcomes. This observable may be vector-valued. The *empirical average* of the observable in a set of $N$ measurements with outcomes $\{k_1, \ldots, k_N\}$ is $\sum_{n=1}^{N} O_{k_n}/N$, equivalent to $\sum_k O_k f_k$.

Probabilities have propositions as arguments (for good definitions of what a proposition is – it isn't a sentence, for example – see Strawson 1964; Copi 1979; Barwise et al. 2003). Johnson's definition remains one of the simplest and most beautiful: "Probability is a magnitude to be attached to any possibly true or possibly false proposition; not, however, to the proposition in and for itself, but in reference to another proposition the truth of which is supposed to be known" (Johnson 1924 Appendix, § 2). See also Hailperin's (1996; 2011) formalization, sadly neglected in the literature. The assumptions or knowledge underlying our probabilities – our "model" – will be generically denoted by $I$, with subscripts denoting specific assumptions. We will sometimes let a quantity stand as abbreviation for a proposition, for example $\boldsymbol{f}$ for "the observed relative frequencies in $N$ measurements are $\boldsymbol{f}$". In such cases the probability symbol will be in lower-case to remind us of our notational sins.

Lest this note become an anthill of indices let's use the following notation: for positive $K$-tuples $\boldsymbol{x} \coloneqq (x_i)$, $\boldsymbol{y} \coloneqq (y_i)$, and number $a$,

$$\begin{aligned} a\boldsymbol{x} &\coloneqq (ax_i), \qquad \boldsymbol{x}/\boldsymbol{y} \coloneqq (x_i/y_i), \qquad \boldsymbol{x}\boldsymbol{y} \coloneqq (x_i y_i), \qquad \boldsymbol{x}^{\boldsymbol{y}} \coloneqq (x_i^{y_i}), \\ \exp \boldsymbol{x} &\coloneqq (\exp x_i), \qquad \ln \boldsymbol{x} \coloneqq (\ln x_i), \qquad \boldsymbol{x}! \coloneqq (x_i!), \\ \sum \boldsymbol{x} &\coloneqq \sum_k x_k, \qquad \prod \boldsymbol{x} \coloneqq \prod_k x_k, \qquad \binom{a}{a\boldsymbol{x}} \coloneqq \frac{a!}{\prod(a\boldsymbol{x})!}. \end{aligned} \quad (1)$$

The symbol $\delta$ indicates a Dirac delta (Lighthill 1964; even better: Egorov 1990; 2001) or a characteristic function (cf. Knuth 1992), depending on the context.

The Shannon entropy $H(\boldsymbol{x}) \coloneqq -\sum \boldsymbol{x} \ln \boldsymbol{x}$, and the relative Shannon entropy or negative discrimination information $H(\boldsymbol{x}; \boldsymbol{y}) \coloneqq -\sum \boldsymbol{x} \ln(\boldsymbol{x}/\boldsymbol{y})$. Let's keep in mind the important properties

$$H(\boldsymbol{x}; \boldsymbol{y}) \leqslant 0, \qquad H(\boldsymbol{x}; \boldsymbol{y}) = 0 \Leftrightarrow \boldsymbol{x} = \boldsymbol{y}. \quad (2)$$

The problem typically addressed by maximum-entropy is this: given that in a large number $N$ of measurements we have observed an average having value in a *convex* set $A$ (which can consist of a single number),

$$\sum Of \in A, \tag{3}$$

what is the probability of having outcome $k$ in an $(N+1)$th measurement? In symbols,

$$\boxed{P\bigl[E_k^{(N+1)}\bigm| \sum Of \in A, I\bigr] = ?} \tag{4}$$

where $I$ denotes our state of knowledge. The maximum-entropy answer (Mead et al. 1984; Fang et al. 1997; Boyd et al. 2009) has the form

$$\frac{r_k \exp(\lambda O_k)}{\sum r \exp(\lambda O)} \tag{5}$$

where $r$ is a reference distribution and $\lambda$ is determined by the constraints in a way that we don't need to specify here. The convexity of $A$ ensures the uniqueness of this answer.

## 2   Maximum-entropy from a model by exchangeability

Let's assume that in our state of knowledge $I_x$ we deem the measurements to be infinitely exchangeable (Bernardo et al. 2000 § 4.2); that is, there can be a potentially unlimited number of them and their indices are irrelevant for our inferences. De Finetti's theorem (1930; 1937; Heath et al. 1976) states that this assumption *forces* us to assign probabilities of this form:

$$\begin{aligned} P\bigl[E_{k_1}^{(1)},\ldots,E_{k_N}^{(N)}\bigm| I_x\bigr] &= \int q_{k_1}\cdots q_{k_N}\, p(q|I_x)\,dq \\ &\equiv \int \Bigl(\prod q^{Nf}\Bigr) p(q|I_x)\,dq, \end{aligned} \tag{6}$$

where the distribution $q$ can be interpreted as the relative frequencies in the long run,[1] and integration is over the $(K-1)$-dimensional simplex (Grünbaum 2003) of such distributions, $\{q \in \mathbf{R}_{\geqslant 0}^K \mid \sum q = 1\}$. The term $p(q|I_x)\,dq$ can be interpreted as the prior probability density of observing the long-run frequencies $q$ in an infinite number of measurements. This probability is not determined by the theorem.

Let's call the expression above an *exchangeability model* (Bernardo et al. 2000 § 4.3).

---

[1] "But this *long run* is a misleading guide to current affairs. *In the long run* we are all dead." (Keynes 2013 § 3.I, p. 65)

We assume that our state of knowledge $I_\text{x}$ is also expressed by a particular prior density for the long-run frequencies:

$$\text{p}(\boldsymbol{q}|\,I_\text{x})\,\text{d}\boldsymbol{q} = \kappa(L,\boldsymbol{r})\binom{L}{L\boldsymbol{q}}\prod r^{L\boldsymbol{q}}\,\text{d}\boldsymbol{q}, \qquad L \geqslant 1, \tag{7}$$

which we can call "multinomial prior" because is a sort of continuous interpolation of the multinomial distribution (Johnson et al. 1996 ch. 35). in the latter each $q_k$ assumes discrete values in $\{0, 1/L, \ldots, 1\}$ and the normalizing constant is unity; for this reason the normalizing constant $\kappa(L, \boldsymbol{r}) \approx L$ in eq. (7). The results that follow also hold for any other prior density that is asymptotically equal to the one above for $L$ large, for example proportional to $\exp[LH(\boldsymbol{q};\boldsymbol{r})]$, which appears in Rodríguez's (1989; 2002) entropic prior and in Skilling's (1989a; 1990) prior for "classical" and "quantified" maximum-entropy.

To find the probability (4) queried by maximum-entropy we need the probability for each possible frequency distribution in the $N$ measurements, which by combinatorial arguments is

$$\text{p}(\boldsymbol{f}|\,I_\text{x}) = \int \binom{N}{N\boldsymbol{f}}\left(\prod q^{N\boldsymbol{f}}\right)\text{p}(\boldsymbol{q}|\,I_\text{x})\,\text{d}\boldsymbol{q}. \tag{8}$$

There are $\binom{N+K-1}{K-1}$ possible frequency distributions (Csiszár et al. 2004).

By marginalization over the subset of frequencies consistent with our data, the probability for the empirical average is

$$\text{P}(\textstyle\sum \boldsymbol{O}\boldsymbol{f} \in A|\,I_\text{x}) = \sum_{\boldsymbol{f}} \delta(\textstyle\sum \boldsymbol{O}\boldsymbol{f} \in A) \int \binom{N}{N\boldsymbol{f}}\left(\prod q^{N\boldsymbol{f}}\right)\text{p}(\boldsymbol{q}|\,I_\text{x})\,\text{d}\boldsymbol{q}. \tag{9}$$

Finally using Bayes's theorem with the probabilities (6)–(9) we find

$$\text{P}\bigl[E_k^{(N+1)}\bigm|\,\textstyle\sum \boldsymbol{O}\boldsymbol{f} \in A, I_\text{x}\bigr] = \frac{\int q_k \sum_{\boldsymbol{f}} \delta(\sum \boldsymbol{O}\boldsymbol{f} \in A)\binom{N}{N\boldsymbol{f}}\left(\prod q^{N\boldsymbol{f}}\right)\text{p}(\boldsymbol{q}|\,I_\text{x})\,\text{d}\boldsymbol{q}}{\int \sum_{\boldsymbol{f}} \delta(\sum \boldsymbol{O}\boldsymbol{f} \in A)\binom{N}{N\boldsymbol{f}}\left(\prod q^{N\boldsymbol{f}}\right)\text{p}(\boldsymbol{q}|\,I_\text{x})\,\text{d}\boldsymbol{q}}, \tag{10}$$

where the density $\text{p}(\boldsymbol{q}|\,I_\text{x})\,\text{d}\boldsymbol{q}$ is specified in eq. (7), even though the formula above holds as well with any other prior density.

I have graphically emphasized this formula because it is the *exact* answer given to the question (4) by a general exchangeability model: it holds for all numbers $K$ of possible outcomes, all numbers $N$ of observations, and all sets $A$ – even *non-convex* ones.

If $N$ and $L$ are large we can use the bounds of the multinomial (Csiszár et al. 1981 Lemma 2.3)

$$\binom{N}{N\boldsymbol{f}} = \epsilon(N,\boldsymbol{f})\exp[N\,H(\boldsymbol{f})], \quad (N+1)^{-K} \leqslant \epsilon(N,\boldsymbol{f}) \leqslant 1, \quad (11)$$

analogously for $\binom{L}{L\boldsymbol{r}}$.

From the bounds above it can be shown that the exact probability expression (10) has the asymptotic form

$$P\bigl[E_k^{(N+1)}\bigm| \sum \boldsymbol{O}\boldsymbol{f} \in A, I_x\bigr] \simeq$$
$$\kappa(N,L,\boldsymbol{r}) \int q_k \sum_{\boldsymbol{f}} \delta(\sum \boldsymbol{O}\boldsymbol{f} \in A) \exp[N H(\boldsymbol{f};\boldsymbol{q}) + L H(\boldsymbol{q};\boldsymbol{r})]\,\mathrm{d}\boldsymbol{q},$$
$$N, L \text{ large.} \quad (12)$$

I prefer the symbol "≃", "is asymptotically equal to" (ISO 2009; IEEE 1993; IUPAC 2007), to the limit symbol "→" because the latter may invite to think about a *sequence*, but no such sequence exists. In each specific problem $N$ has one, fixed, possibly unknown value, and cannot be increased at will. The symbol "≃" says that the right side differs from the left side by an error that may be negligible. It is our duty to check whether this error is really negligible for our purposes.

The asymptotic expression above shows an interesting interplay of two relative entropies. The two exponential terms give rise to two Dirac deltas. The delta in $\boldsymbol{f}$ requires some mathematical care owing to the discreteness of this quantity; see Csiszár (1984; 1985). In particular, if $N < K$ the discrete set of $\binom{N+K-1}{K-1}$ possible frequency distributions lies within the $(N-1)$-dimensional facets of the $(K-1)$-dimensional simplex of distributions $\boldsymbol{q}$; it does not "fill" the simplex. In this case the frequency sum $\sum_{\boldsymbol{f}}$ cannot be meaningfully approximated by an integral. The approximations below are valid if the number $N$ of observations is much larger than the number $K$ of possible outcomes.

If $L/N$ is also large, taking limits in the proper order gives

$$P\bigl[E_k^{(N+1)}\bigm| \sum \boldsymbol{O}\boldsymbol{f} \in A, I_x\bigr] \simeq r_k, \qquad N, L, L/N \text{ large.} \quad (13)$$

Note how the data about the average (3) are practically discarded in this $(L/N)$-large case. Compare with Skilling's remark that the parameter $L$ (his $\alpha$) shouldn't be "particularly large" (cf. Skilling 1998 p. 2).

The asymptotic case that interests us is $N/L$ large: the exponential in $N$ dominates the integral of eq. (12), which becomes

$$\kappa(L, \boldsymbol{r}) \sum_{\boldsymbol{f}} f_k\, \delta[\sum \boldsymbol{Of} \in A]\, \exp[LH(\boldsymbol{f};\boldsymbol{r})] \simeq \arg\sup_{\boldsymbol{f}}^{\sum \boldsymbol{Of} \in A} H(\boldsymbol{f};\boldsymbol{r}), \qquad (14)$$

so that, finally,

$$\mathrm{P}\bigl[E_k^{(N+1)}\bigm|\, \sum \boldsymbol{Of} \in A, I_{\mathrm{x}}\bigr] \simeq f_k^*, \qquad N, L, L/N \text{ large},$$
$$\text{with } \boldsymbol{f}^* \text{ maximizing } H(\boldsymbol{f};\boldsymbol{r}) \text{ under constraints } \sum \boldsymbol{Of} \in A, \qquad (15)$$

which is the maximum-entropy recipe, giving the distribution (5).

## 3 Maximum-entropy from a model by sufficiency

Consider the following assumption or working hypothesis, denoted $I_{\mathrm{s}}$: To predict the outcome of an $(N+1)$th measurement given knowledge of the outcomes of $N$ measurements, all we need to know is the average $\sum \boldsymbol{Of}$ of an observable $\boldsymbol{O}$ in those $N$ measurements, no matter the value of $N$. In other words, any data about known measurements, besides the empirical average of $\boldsymbol{O}$, is irrelevant for our prediction. The average $\sum \boldsymbol{Of}$ is then called a *minimal sufficient statistics* (Bernardo et al. 2000 § 4.5; Lindley 2008 § 5.5). In symbols,

$$\mathrm{P}\bigl[E_k^{(N+1)}\bigm|\, E_{k_1}^{(1)}, \dots, E_{k_N}^{(N)}, I_{\mathrm{s}}\bigr] = \mathrm{p}\bigl[E_k^{(N+1)}\bigm|\, \sum \boldsymbol{Of}, N, I_{\mathrm{s}}\bigr]. \qquad (16)$$

Note that the data $\{E_{k_n}^{(n)}\}$ determine the data $\{\sum \boldsymbol{Of}, N\}$ but not vice versa, so some data have effectively been discarded in the conditional.

The Koopman-Pitman-Darmois theorem (Koopman 1936; Pitman 1936; Darmois 1935; see also later analyses: Hipp 1974; Andersen 1970; Denny 1967; Fraser 1963; Barankin et al. 1963) states that this assumption forces us to assign probabilities of this form:

$$\mathrm{P}\bigl[E_{k_1}^{(1)}, \dots, E_{k_N}^{(N)}\bigm|\, I_{\mathrm{s}}\bigr] = \int \mathrm{p}(k_1|\lambda, \boldsymbol{r}, I_{\mathrm{s}}) \cdots \mathrm{p}(k_N|\lambda, \boldsymbol{r}, I_{\mathrm{s}})\, \mathrm{p}(\lambda|I_{\mathrm{s}})\, \mathrm{d}\lambda,$$
$$\equiv \int \Bigl[\prod \mathrm{p}(\boldsymbol{k}|\lambda, \boldsymbol{r}, I_{\mathrm{s}})^{Nf}\Bigr] \mathrm{p}(\lambda|I_{\mathrm{s}})\, \mathrm{d}\lambda,$$
$$(17\mathrm{a})$$

$$\text{with}\quad \mathrm{p}(\boldsymbol{k}|\lambda, \boldsymbol{r}, I_{\mathrm{s}}) \coloneqq \boldsymbol{r}\,\frac{\exp(\lambda \boldsymbol{O})}{Z(\lambda)}, \quad Z(\lambda) \coloneqq \sum \boldsymbol{r}\exp(\lambda \boldsymbol{O}), \qquad (17\mathrm{b})$$

and we have defined $p(\mathbf{k}|\dots) \coloneqq \bigl(p(1|\dots),\dots,p(K|\dots)\bigr)$. The integration of the parameter $\lambda$ is over $\mathbf{R}^M$, with $M$ the dimension of the vector-valued observable $\mathbf{O}$, and $\mathbf{r}$ is a $K$-dimensional distribution. Neither $\mathbf{r}$ or the distribution $p(\lambda|I_s)$ are determined by the theorem.

Let's call the expression above a *sufficiency model* (Bernardo et al. 2000 § 4.5). A sufficiency model can be viewed as a mixture, with weight density $p(\lambda|I_s)\,\mathrm{d}\lambda$, of distributions having maximum-entropy form (5) with multipliers $\lambda$.

To find the probability (4) we calculate, as in the previous section, the probabilities for the frequencies:

$$p(f|I_s) = \int \binom{N}{Nf} \left[\prod p(\mathbf{k}|\lambda,\mathbf{r},I_s)^{Nf}\right] p(\lambda|I_s)\,\mathrm{d}\lambda, \qquad (18)$$

and for the empirical average by marginalization:

$$P\bigl(\textstyle\sum \mathbf{O}f \in A\big|\, I_s\bigr) =$$
$$\sum_f \delta(\textstyle\sum \mathbf{O}f \in A) \int \binom{N}{Nf} \left[\prod p(\mathbf{k}|\lambda,\mathbf{r},I_s)^{Nf}\right] p(\lambda|I_s)\,\mathrm{d}\lambda. \quad (19)$$

From these using Bayes's theorem we finally find

$$P\bigl[E_k^{(N+1)}\big|\, \textstyle\sum \mathbf{O}f \in A, I_s\bigr] =$$
$$\frac{\int p(\mathbf{k}|\lambda,\mathbf{r},I_s) \sum_f \delta(\sum \mathbf{O}f \in A)\binom{N}{Nf}\left[\prod p(\mathbf{k}|\lambda,\mathbf{r},I_s)^{Nf}\right] p(\lambda|I_s)\,\mathrm{d}\lambda}{\int \sum_f \delta(\sum \mathbf{O}f \in A)\binom{N}{Nf}\left[\prod p(\mathbf{k}|\lambda,\mathbf{r},I_s)^{Nf}\right] p(\lambda|I_s)\,\mathrm{d}\lambda}.$$
$$(20)$$

This is the *exact* answer given to the maximum-entropy question by a sufficiency model *if the constraints used in maximum-entropy are considered to be a sufficient statistics*. This proviso has serious consequences discussed in § 5.2. The expression above holds for all $N$ and all sets $A$, even *non-convex* ones.

The asymptotic analysis for large $N$ uses again the multinomial's bounds (11). We find

$$P\bigl[E_k^{(N+1)}\big|\, \textstyle\sum \mathbf{O}f \in A, I_s\bigr] \simeq \kappa(N,\mathbf{r})\int p(\mathbf{k}|\lambda,\mathbf{r},I_s)\,\times$$
$$\sum_f \delta(\textstyle\sum \mathbf{O}f \in A)\exp\bigl\{NH\bigl[f;p(\mathbf{k}|\lambda,\mathbf{r},I_s)\bigr]\bigr\}\,p(\lambda|I_s)\,\mathrm{d}\lambda,$$
$$N \text{ large.} \quad (21)$$

A rigorous analysis of this limit can be done using "information projections" (Csiszár 1984; 1985); here is a heuristic summary. Consider the sum in $f$ for fixed $\lambda$. We have two cases. (1) If $\lambda$ is such that $\sum \boldsymbol{O}\, \mathrm{p}(\boldsymbol{k}|\lambda, \boldsymbol{r}, I_\mathrm{s}) \in A$, there exists a unique $f$ in the sum for which the relative entropy in the exponential reaches its maximum, zero, making the exponential unity. For all other $f$ the relative entropy is negative and the exponential asymptotically vanishes for large $N$. The integral therefore doesn't vanish asymptotically. (2) If $\lambda$ is such that $\mathrm{p}(\boldsymbol{k}|\lambda, \boldsymbol{r}, I_\mathrm{s})$ doesn't satisfy the constraints, the relative entropy in the exponential will be negative for all $f$ in the sum, making the exponential asymptotically vanish for all $f$. The integral therefore vanishes asymptotically. The distinction between these two cases actually requires mathematical care owing to the discreteness of the sum. The $f$ sum then acts as a delta or characteristic function (depending on whether $A$ has measure zero or not):

$$\sum_f \delta(\sum \boldsymbol{O} f \in A)\, \exp\{N H[f; \mathrm{p}(\boldsymbol{k}|\lambda, \boldsymbol{r}, I_\mathrm{s})]\} \simeq \delta[\sum \boldsymbol{O}\, \mathrm{p}(\boldsymbol{k}|\lambda, \boldsymbol{r}, I_\mathrm{s}) \in A]. \tag{22}$$

Thus asymptotically we have, using the explicit expression (17b) for $\mathrm{p}(\boldsymbol{k}|\lambda, \boldsymbol{r}, I_\mathrm{s})$:

$$\mathrm{P}[E_k^{(N+1)} | \sum \boldsymbol{O} f \in A, I_\mathrm{s}] \simeq \\ \int \delta\!\left[\sum \boldsymbol{O}\, \boldsymbol{r}\, \frac{\exp(\lambda \boldsymbol{O})}{Z(\lambda)} \in A\right] r_k \frac{\exp(\lambda O_k)}{Z(\lambda)}\, \mathrm{p}(\lambda|I_\mathrm{s})\, \mathrm{d}\lambda, \quad N \text{ large.} \tag{23}$$

This result can also be found first integrating $\lambda$ and then summing $f$, using a heuristic argument similar to the one above. This is a mixture, with weight density $\mathrm{p}(\lambda|I_\mathrm{s})\,\mathrm{d}\lambda$, of maximum-relative-entropy distributions $f^*$ that satisfy the individual constraints $\sum \boldsymbol{O} f^* = a$, $a \in A$. The final distribution thus differs from the maximum-entropy one if the set $A$ is not a singleton: maximum-entropy would pick up only one distribution. But if the constraint set is a singleton, $A = \{a\}$, we do obtain the same answer (5) as the maximum-entropy recipe:

$$\mathrm{P}[E_k^{(N+1)} | \sum \boldsymbol{O} f = a, I_\mathrm{x}] \simeq f_k^*, \qquad N \text{ large},$$
$$\text{with } f^* \text{ maximizing } H(f; \boldsymbol{r}) \text{ under constraints } \sum \boldsymbol{O} f = a. \tag{24}$$

## 4  Heuristic explanation of both asymptotic approximations

First of all let's note that both the exchangeability (6) and sufficiency (17) models have the parametric form

$$\mathrm{P}\bigl[E^{(1)}_{k_1},\ldots,E^{(N)}_{k_N}\bigm|I\bigr] = \int \mathrm{p}(k_1|\,\nu,I)\cdots\mathrm{p}(k_N|\,\nu,I)\,\mathrm{p}(\nu|\,I)\,\mathrm{d}\nu \\ \equiv \int \Bigl[\prod \mathrm{p}(\bm{k}|\,\nu,I)^{Nf}\Bigr]\mathrm{p}(\nu|\,I)\,\mathrm{d}\nu. \quad(25)$$

The final probability distribution $\bm{p}$ for the $K$ outcomes of the $(N+1)$th measurement belongs to a $(K-1)$-dimensional simplex $\{\bm{p}\in\mathbf{R}^K_{\geqslant 0}\mid\sum\bm{p}=1\}$. The expression above first selects, within this simplex, a family of distributions $\{\mathrm{p}(\bm{k}|\,\nu,I)\}$ parametrized by $\nu$; then it delivers the distribution $\bm{p}$ as a mixture of the distributions of this family, with weight density $\mathrm{p}(\nu|\,I)\,\mathrm{d}\nu$. In the exchangeability model this family is actually the whole simplex (that's why it's sometimes called a "non-parametric" model). In the sufficiency model it is an *exponential family* (Bernardo et al. 2000 § 4.5.3; Barndorff-Nielsen 2014).

When we conditionalize on data $D$, the weight density is determined by the mutual modulation of two weights: that of the probability of the data $\mathrm{p}(D|\,\nu,I)$ and the initial weight $\mathrm{p}(\nu|\,I)$. Pictorially, if $K=3$:

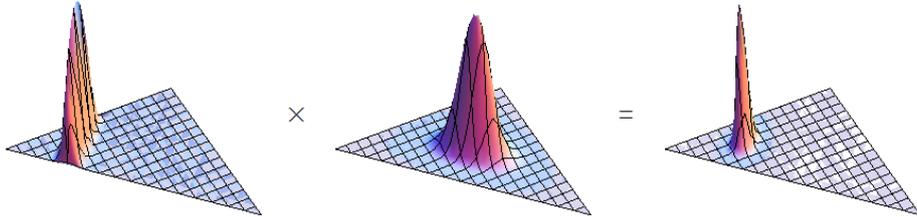

$$\mathrm{p}(D|\,\nu,I) \quad \times \quad \mathrm{p}(\nu|\,I) \quad = \quad \kappa\,\mathrm{p}(\nu|\,D,I) \quad(26)$$

the final $\bm{p}$ is given by the mixture with the weight density $\mathrm{p}(\nu|\,D,I)\,\mathrm{d}\nu$ ensuing from this modulation. The mathematical expression of the data weight $\mathrm{p}(D|\,\nu,I)$ is typically exponentiated to the number of measurements $N$ from which the data originate; compare with eqs (19), (20). If $N$ is large this weight is very peaked on the subset of distributions that give highest probability to the data, that is, that have expectations very close to the empirical averages. It effectively restricts the second weight $\mathrm{p}(\nu|\,I)\,\mathrm{d}\nu$ to such "data subset". In our case the data subset consists of all distributions satisfying the constraints.

The mechanism described so far is common to the exchangeability and the sufficiency model. Their difference lies in how they choose the final distribution from the data subset.

In the exchangeability model (6) the choice is made by the weight density $p(\nu|I)\,d\nu$, i.e. the multinomial prior (7). It is extremely peaked owing to the large parameter $L$, and its level curves are isentropics. Once it's restricted to the data subset by the data weight $p(D|\nu, I)$, it gives highest weight to the distribution $p$ lying on the highest isentropic curve, which is unique if the data subset is convex; compare with fig.-eq. (26). Hence this is a maximum-entropy distribution satisfying the data constraints. For this mechanism to work it's necessary that the dominance of the data weight comes first, and the dominance of the multinomial prior comes second. This is the reason why the correct asymptotic limit (15) has $N$, $L$, and $N/L$ large.

In the sufficiency model (17) the choice is made by the family of distributions $\{p(k|\nu, I)\}_\nu$. These distributions have by construction a maximum-entropy form for the particular observable $O$. This family intersects the data subset in only one point if the constraint has the form $\sum Of = a$. This point is therefore the maximum-entropy distribution satisfying the data constraints.

The mechanism above also explains why these two models still work if the data subset is non-convex and touches the highest isentropics (exchangeability model) or the exponential family (sufficiency model) in multiple points, bringing the maximum-entropy recipe to an impasse. The final distribution will simply be an equal mixture of such tangency points; it may well lie outside of the data subset.

## 5  Discussion

**5.1  Prediction or retrodiction?**  An essential aspect of the maximum-entropy method is surprisingly often disregarded in the literature. If we have data from $N$ measurements, we can ask two questions:

**"Prediction":** what is the outcome of a *further* similar measurement?

**"Retrodiction":** what is the outcome of the first of the $N$ measurements?

Note that despite the literal meaning of these terms *the distinction is not between future and past, but between unknown and partially known*.

It's rarely made clear whether the maximum-entropy probabilities refer to the first or to the second question. Yet these two questions are fundamentally different; their answers rely on very different principles.

To answer the first question we can – but need not – fully rely on symmetry principles in the discrete case. It is a matter of combinatorics and equal probabilities; a drawing-from-an-urn problem. Most derivations of the maximum-entropy method (e.g. Jaynes 1963; Shore et al. 1980; van Campenhout et al. 1981; Csiszár 1985) address this question only, as often betrayed by the presence of "$p(x_1)$" or similar expressions in their final formulae.

To answer the second question, symmetry and combinatorics alone are no use: additional principles are needed. This is the profound philosophical question of *induction*, with its ocean of literature; my favourite sample are the classic Hume (1896 book I, § III.VI), Johnson (1922 esp. chs VIII ff; 1924 Appendix; 1932), de Finetti (1937; 1959), Jeffreys (1955; 1973 ch. I; 2003 § 1.0), Jaynes (2003 § 9.4). De Finetti, foreshadowed by Johnson, was probably the one who expressed most strongly, and explained brilliantly, that the probability calculus does not and cannot *explain* or *justify* our inductive reasoning; it only *expresses* it in a quantitative way. This shift in perspective was very much like Galilei's shift from *why* to *how* in the study of physical phenomena.[2] We do inductive inferences in many different ways (Jaynes 2003 § 9.4). The notion of exchangeability (de Finetti 1937; Johnson 1924 Appendix; 1932) captures one of the most intuitive and expresses it mathematically.

The calculations of the previous sections and the final probabilities (10), (20) for our two models pertain the predictive question, as clear from the $E^{(N+1)}$ in their arguments. The two models can also be used to answer the retrodictive question. The resulting formulae are different; they can again be found applying the rules of the probability calculus and Bayes's theorem. The retrodictive formula for the exchangeability model is (proof

---

[2]"According to credible traditions it was in the sixteenth century, an age of very intense spiritual emotions, that people gradually ceased trying, as they had been trying all through two thousand years of religious and philosophic speculation, to penetrate into the secrets of Nature, and instead contented themselves, in a way that can only be called superficial, with investigations of its surface. The great Galileo, who is always the first to be mentioned in this connection, did away with the problem, for instance, of the intrinsic reasons why Nature abhors a vacuum, so that it will cause a falling body to enter into and occupy space after space until it finally comes to rest on solid ground, and contented himself with a much more general observation: he simply established the speed at which such a body falls, what course it takes, what time it takes, and what its rate of acceleration is. The Catholic Church made a grave mistake in threatening this man with death and forcing him to recant, instead of exterminating him without more ado." (Musil 1979 vol. 1, ch. 72)

in Porta Mana 2009 § B):

$$\mathrm{P}\bigl[E_k^{(n)}\bigm| \textstyle\sum \boldsymbol{O}\boldsymbol{f} \in A, I_\mathrm{x}\bigr] = \frac{\int \sum_f f_k\, \delta(\sum \boldsymbol{O}\boldsymbol{f} \in A) \binom{N}{N\boldsymbol{f}} \bigl(\prod \boldsymbol{q}^{N\boldsymbol{f}}\bigr) \mathrm{p}(\boldsymbol{q}|\,I_\mathrm{x})}{\int \sum_f \delta(\sum \boldsymbol{O}\boldsymbol{f} \in A) \binom{N}{N\boldsymbol{f}} \bigl(\prod \boldsymbol{q}^{N\boldsymbol{f}}\bigr) \mathrm{p}(\boldsymbol{q}|\,I_\mathrm{x})\,\mathrm{d}\boldsymbol{q}}\,\mathrm{d}\boldsymbol{q},$$
$$n \in \{1,\dots,N\}. \quad (27)$$

Graphically it differs from the predictive one (10) only in the replacement of $q_k$ by $f_k$. An analogous replacement appears in the retrodictive formula for the sufficiency model. But this graphically simple replacement leads to a mechanism very different from the one of § 4 in delivering the final probability: it's a mixture on the data subset rather than on the whole simplex. Predictive and retrodictive probabilities can therefore be very different for small $N$. See for example figs 1 and 2 below and their accompanying discussion.

This means that the goodness of the maximum-entropy distribution as an approximation of our two models can depend on whether we are asking a *predictive* or a *retrodictive* question. This fact is very important in every application.

**5.2  Which of the two models is preferable?**  A maximum-entropy distribution can be seen as an approximation of the distribution obtained from the exchangeability model or the sufficiency one (*repetita iuvant*). The two inferential models are not equivalent though, and there are reasons to prefer the exchangeability one – despite the frequent association, in the literature, of maximum-entropy with exponential families. The most important and quite serious difference is this:

Suppose that we have used either model to assign a predictive distribution conditional on the empirical average $a$ of the observable $\boldsymbol{O}$, obtained from $N$ measurements. If $N$ is large the distributions obtained from either model will be approximately equal, and equal to the maximum-entropy one. Now someone gives us a new empirical average $a'$ of a different observable $\boldsymbol{O}'$, obtained from the *same $N$* measurements. This observable turns out to be complementary to the previous one, in the sense that in general from knowing the value of $\sum \boldsymbol{O}\boldsymbol{f}$ we cannot deduce the value of $\sum \boldsymbol{O}'\boldsymbol{f}$, and vice versa. These new data therefore reveal more about the outcomes of our $N$ measurements and of possible further measurements.

The new empirical average $a'$ can be incorporated in the exchangeability model; the resulting predictive and retrodictive distributions conditional on $(a', a)$ will be numerically different from the ones conditional on $a$ only.

They will be approximated by a maximum-entropy one based on the old *and* new constraints.

If we incorporate the new average in the sufficiency model, however, *the resulting* predictive *conditional distribution will be unchanged: knowledge of the new data has no effect in the prediction of new measurements*. The reason is simple: the sufficiency model expresses by construction that the average of the old observable $O$ is all we need for our inferences about further measurements. Any other observable is irrelevant. The new average automatically drops out under predictive conditioning. The only way to obtain a different predictive conditional distribution would be to *discard* the sufficiency model based on $O$, and use a new one based on $(O, O')$. But that would be cheating!

This shows how dramatically absolute and categorical the assumption of the existence of a sufficient statistics is. The difficulty above doesn't happen for the retrodictive distribution; the proof is left as an exercise for you.

Since the maximum-entropy method is meant to always employ new constraints, we deduce that it's more correct to interpret it as an approximation of the exchangeability model than of the sufficiency model.

**5.3 How good is the maximum-entropy approximation?** How does maximum-entropy compare with the exchangeability model (6) with multinomial prior (7) away from the asymptotic approximation?

Their distributions are compared in the classic example of dice rolling in figs 1 and 2 for empirical averages of 5 and 6 (see Porta Mana 2009 for the calculations). The maximum-entropy distribution (red) is at the top; the distribution of the exchangeability model with $L = 1$ (blue) and $L = 50$ (bluish purple) is shown underneath for the cases $N = 2$, $N = 12$, $N = \infty$, and for the retrodiction of an "old roll" $E_k^{(n)}$, $n \in \{1, \ldots, N\}$, and the prediction of a "new roll" $E_k^{(N+1)}$. The charts also report the Shannon entropies $H$ of the distributions.

The exchangeability model gives very reasonable and even "logical" probabilities for small $N$. For example, if you obtain an average of 5 in two rolls, it's impossible that either of them was ⚀ – unless, of course, you own a six-sided die with nine pips on one face. The exchangeability model logically gives zero probability in this case (fig. 1 bottom left). Maximum-entropy gives an erroneous non-zero probability. And having obtained an average of 5 or 6 in two rolls, would you really give a much higher probability to ⚁ or ⚄ for a *third* roll? I'd still give 1/6. The exchangeability

$H = 1.370$

| ⚀ | ⚁ | ⚂ | ⚃ | ⚄ | ⚅ |
|---|---|---|---|---|---|
| 2.1 | 3.9 | 7.2 | 13.6 | 25.5 | 47.8 |

maximum-entropy distribution, $a = 5$

retrodictive                                              predictive

$H_{L=1} = 1.347$
$H_{L=50} = 1.367$

P(old roll| $N = \infty$, $a = 5$, $I_{xL=1 \text{ or } L=50}$)

$H_{L=1} = 1.347$
$H_{L=50} = 1.367$

P(new roll| $N = \infty$, $a = 5$, $I_{xL=1 \text{ or } L=50}$)

$H_{L=1} = 1.363$
$H_{L=50} = 1.366$

P(old roll| $N = 12$, $a = 5$, $I_{xL=1 \text{ or } L=50}$)

$H_{L=1} = 1.609$
$H_{L=50} = 1.777$

P(new roll| $N = 12$, $a = 5$, $I_{xL=1 \text{ or } L=50}$)

$H_{L=1} = 1.045$
$H_{L=50} = 1.097$

P(old roll| $N = 2$, $a = 5$, $I_{xL=1 \text{ or } L=50}$)

$H_{L=1} = 1.756$
$H_{L=50} = 1.791$

P(new roll| $N = 2$, $a = 5$, $I_{xL=1 \text{ or } L=50}$)

Figure 1   Maximum-entropy and exchangeability model, empirical average $a = 5$

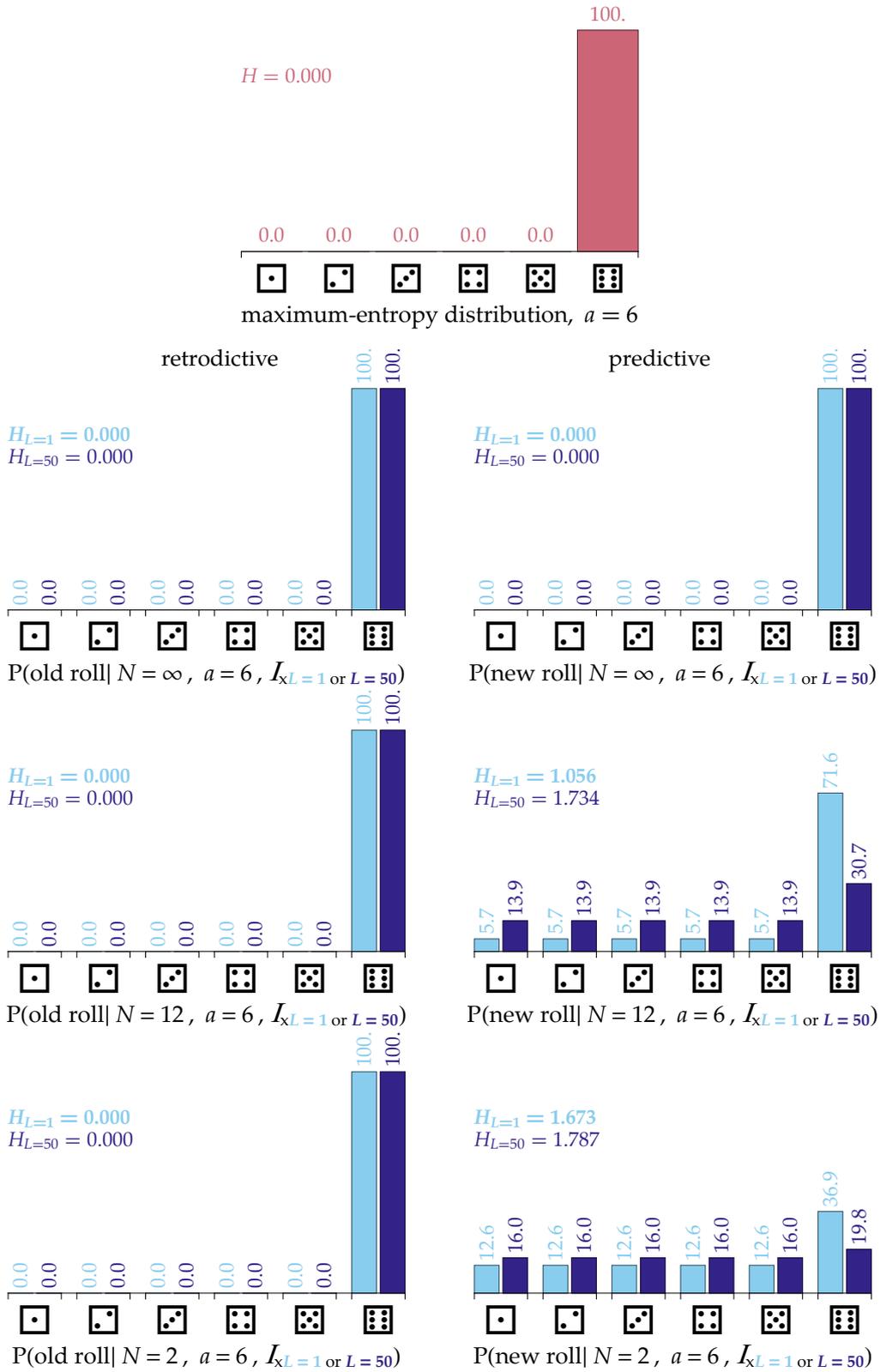

Figure 2   Maximum-entropy and exchangeability model, empirical average $a = 6$

model reasonably gives an almost uniform distribution, especially for large $L$ (both figures bottom right). The maximum-entropy distribution is unreasonably biased towards high values. If we observe a high average in twelve rolls we start to suspect that the die/dice or the roll technique are biased. The exchangeability model expresses this bias, but more conservatively than maximum-entropy.

In fact the predictive exchangeability-model distribution can have *higher* entropy than the maximum-entropy one! This happens because, when $N$ is small compared to $L$, the maximum-entropy prescription "what you've seen in $N$ measurements  =  what you should expect in an $(N+1)$th measurement" is silly (MacKay 2003 Exercise 22.13). The exchangeability model intelligently doesn't respect this prescription strictly, if $N$ isn't large.[3] See Porta Mana (2009) for comparisons under other values of the empirical average and of the number of measurements.

When is $N$ large enough for the prescription to become reasonable? In other words, when is maximum-entropy a good approximation of the exchangeability model with multinomial prior? The answer depends on the interplay among the number of measurements $N$, the number of possible outcomes $K$, the parameter $L$, the reference distribution $r$, and the value $a$ (or range $A$) of the observed average. The first three ingredients determine the maximum heights of the densities involved in the integral and sum of eq. (10); the last three ingredients determine the size of the effective integration and sum region relative to the integration simplex, and the distance between the peaks of the data weights and the prior weights of fig.-eq. (26). All five ingredients determine how good are the delta approximations in the integral and sum of eq. (10). We saw in § 2, p. 5, after eq. (12), that $N$ needs to be much larger than $K$ for the integral and delta approximations of the frequency sum to be meaningful. Maximum-entropy approximations are not meaningful if the number of possible outcomes is much larger than the number of observations.

It would be very useful to have explicit estimates of the maximum-entropy-approximation error as a function of the four quantities above. I hope to analyse them in a future note, and promise it would be a shorter note.

### 5.4  Is this a "derivation" of maximum-entropy?

The heuristic explanation of § 4 shows that the maximum-entropy distributions appear asymptotically owing to our specific choices of a multinomial prior in the exchangeability model, and of an exponential family with observable $O$ in the sufficiency model. They are therefore not derived only from first principles or from some sort of universal limit. This is why I don't call the asymptotic analysis discussed in this note a "derivation" of the maximum-entropy "principle". In my opinion this analysis shows that it is not a principle at all.

---

[3]"Obedience is no longer a virtue." (Milani 1965)

The information-theoretic arguments – or should we say incentives – behind the standard maximum-entropy recipe can be lifted to a *meta*[4] level and used for priors asymptotically equivalent to the multinomial prior (7), as done by Rodríguez (1989; 2002) for the entropic prior (see also Skilling 1989a; 1990). Such arguments don't determine the parameters $L$ and $r$, though. They seem to be prone to an infinite regress; Jaynes was aware of this (Jaynes 2003 § 11.1, p. 344).

It would be useful if the multinomial or entropic priors could be uniquely determined by intuitive inferential assumptions, as for example is the case with the Johnson-Dirichlet prior, proportional to $q^L \, \mathrm{d}q$: this prior *must* be used if we believe (denote this by $I_J$) that the frequencies of other outcomes are irrelevant for predicting a particular one:

$$\mathrm{p}\big[E_k^{(N+1)}\big|\, f, N, I_J\big] = \mathrm{p}\big[E_k^{(N+1)}\big|\, f_k, N, I_J\big], \qquad k \in \{1, \dots, K\}, \qquad (28)$$

a condition called "sufficientness" (Johnson 1924; 1932; Good 1965 ch. 4; Zabell 1982; Jaynes 1996). Asymptotically it leads to a maximum-entropy distribution with Burg's (1975) entropy $\sum \ln x$ (see Jaynes 1996; Porta Mana 2009).

But, after all, the logical calculus doesn't tell us which truths to choose at the beginning of a logical deduction. Why should the probability calculus tell us which probabilities to choose at the beginning of a probabilistic induction?

**5.5 Conclusion** Interpreting the maximum-entropy method as an approximation of the exchangeable model (6) with multinomial prior (7) has many advantages:

- it clears up the meaning of the "expectation = average" prescription of the maximum-entropy method;
- it identifies the range of validity of such prescription;
- it quantifies the error of the maximum-entropy approximation;

---

[4]"This is an expression used to hide the absence of any mathematical idea [...]. Personally, I never use this expression in front of children." (Girard 2001 p. 446)

- it gives a more sensible solution when this approximation doesn't hold;
- it clearly differentiates between prediction and retrodiction;
- it can be backed up by information-theoretic incentives (Rodríguez 1989; 2002) if you're into those.

Disadvantages:

- It can't be used to answer the question "Where did the cat go?". But this question lies forever beyond the reach of the probability calculus.

That's all (Hanshaw 1928).

## Thanks

Porta Mana — Maximum-entropy from the probability calculusBurg, J. P. (1975): *Maximum entropy spectral analysis*. PhD thesis. (Stanford University, Stanford). http://sepwww.stanford.edu/data/media/public/oldreports/sep06/.

Copi, I. M. (1979): *Symbolic Logic*, 5th ed. (Macmillan, New York). First publ. 1954.

Csiszár, I. (1984): *Sanov property, generalized I-projection and a conditional limit theorem*. Ann. Prob. **12**[3], 768–793.

— (1985): *An extended maximum entropy principle and a Bayesian justification*. In: (Bernardo, DeGroot, Lindley, Smith 1985), 83–93. With discussion and reply (Barnard, Jaynes, Seidenfeld, Polasek, Csiszár 1985).

Csiszár, I., Körner, J. (1981): *Information Theory: Coding Theorems for Discrete Memoryless Systems*. (Academic Press, New York). Second ed. (Csiszár, Körner 2011).

— (2011): *Information Theory: Coding Theorems for Discrete Memoryless Systems*, 2nd ed. (Cambridge University Press, Cambridge). First publ. 1981.

Csiszár, I., Shields, P. C. (2004): *Information theory and statistics: a tutorial*. Foundations and Trends in Communications and Information Theory **1**[4], 417–528. http://www.renyi.hu/~csiszar/.

Curien, P.-L. (2001): *Preface to Locus solum*. Math. Struct. in Comp. Science **11**[3], 299–300. See also (Girard 2001).

Darmois, G. (1935): *Sur les lois de probabilité à estimation exhaustive*. Comptes rendus hebdomadaires des séances de l'Académie des sciences **200**, 1265–1266.

de Finetti, B. (1930): *Funzione caratteristica di un fenomeno aleatorio*. Atti Accad. Lincei: Sc. Fis. Mat. Nat. **IV**[5], 86–133. http://www.brunodefinetti.it/Opere.htm.

— (1937): *La prévision : ses lois logiques, ses sources subjectives*. Ann. Inst. Henri Poincaré **7**[1], 1–68. Transl. as (de Finetti 1964).

— (1959): *La probabilità e la statistica nei rapporti con l'induzione, secondo i diversi punti di vista*. In: (de Finetti 2011), 1–115. Transl. as (de Finetti 1972a).

— (1964): *Foresight: its logical laws, its subjective sources*. In: (Kyburg, Smokler 1980), 53–118. Transl. of (de Finetti 1937) by Henry E. Kyburg, Jr.

— (1972a): *Probability, statistics and induction: their relationship according to the various points of view*. In: (de Finetti 1972b), ch. 9, 147–227. Transl. of (de Finetti 1959).

— (1972b): *Probability, Induction and Statistics: The art of guessing*. (Wiley, London).

— ed. (2011): *Induzione e statistica*, reprint. (Springer, Berlin). First publ. 1959.

Demidov, A. S. (2001): *Generalized Functions in Mathematical Physics: Main Ideas and Concepts*. (Nova Science, Huntington, USA). With an addition by Yu. V. Egorov.

Denny, J. L. (1967): *Sufficient conditions for a family of probabilities to be exponential*. Proc. Natl. Acad. Sci. (USA) **57**[5], 1184–1187.

Diaconis, P., Freedman, D. (1981): *Partial exchangeability and sufficiency*. In: (Ghosh, Roy 1981), 205–236. Also publ. 1982 as technical report https://www.stat.berkeley.edu/~aldous/206-Exch/Papers/diaconis_freedman_PES.pdf, http://statweb.stanford.edu/~cgates/PERSI/year.html.

Egorov, Yu. V. (1990): *A contribution to the theory of generalized functions*. Russ. Math. Surveys (Uspekhi Mat. Nauk) **45**[5], 1–49.

— (2001): *A new approach to the theory of generalized functions*. In: (Demidov 2001), 117–123.

Erickson, G. J., Rychert, J. T., Smith, C. R., eds. (1998): *Maximum Entropy and Bayesian Methods*. (Springer, Dordrecht).

Fang, S.-C., Rajasekera, J. R., Tsao, H.-S. J. (1997): *Entropy Optimization and Mathematical Programming*, reprint. (Springer, New York).

Ford, K. W., ed. (1963): *Statistical Physics*. (Benjamin, New York).
19

Porta Mana · Maximum-entropy from the probability calculus
Pitman, E. J. G. (1936): *Sufficient statistics and intrinsic accuracy*. Math. Proc. Camb. Phil. Soc. **32**[4], 567–579.

Porta Mana, P. G. L. (2009): *On the relation between plausibility logic and the maximum-entropy principle: a numerical study*. arXiv:0911.2197. Presented as invited talk at the 31st International Workshop on Bayesian Inference and Maximum Entropy Methods in Science and Engineering "MaxEnt 2011", Waterloo, Canada.

Rodríguez, C. C. (1989): *The metrics induced by the Kullback number*. In: (Skilling 1989b), 415–422.

— (2002): *Entropic priors for discrete probabilistic networks and for mixtures of Gaussians models*. Am. Inst. Phys. Conf. Proc. **617**, 410–432. arXiv:physics/0201016.

Shore, J. E., Johnson, R. W. (1980): *Axiomatic derivation of the principle of maximum entropy and the principle of minimum cross-entropy*. IEEE Trans. Inform. Theor. **IT-26**[1], 26–37. See also comments and correction (Shore, Johnson 1983).

— (1983): *Comments on and correction to "axiomatic derivation of the principle of maximum entropy and the principle of minimum cross-entropy"*. IEEE Trans. Inform. Theor. **IT-29**[6], 942–943.

Sivia, D. S. (2006): *Data Analysis: A Bayesian Tutorial*, 2nd ed. (Oxford University Press, Oxford). Written with J. Skilling. First publ. 1996.

Skilling, J. (1989a): *Classic maximum entropy*. In: (Skilling 1989b), 45–52.

— ed. (1989b): *Maximum Entropy and Bayesian Methods: Cambridge, England, 1988*. (Kluwer, Dordrecht).

— (1990): *Quantified maximum entropy*. In: (Fougère 1990), 341–350.

— (1998): *Massive inference and maximum entropy*. In: (Erickson, Rychert, Smith 1998), 1–14. http://www.maxent.co.uk/documents/massinf.pdf.

Strawson, P. F. (1964): *Introduction to Logical Theory*. (Methuen, London). First publ. 1952.

van Campenhout, J. M., Cover, T. M. (1981): *Maximum entropy and conditional probability*. IEEE Trans. Inform. Theor. **IT-27**[4], 483–489.

Zabell, S. L. (1982): *W. E. Johnson's "sufficientness" postulate*. Ann. Stat. **10**[4], 1090–1099. Repr. in (Zabell 2005 pp. 84–95).

— (2005): *Symmetry and Its Discontents: Essays on the History of Inductive Probability*. (Cambridge University Press, Cambridge).
22